\documentclass[twocolumn,superscriptaddress,aps,floatfix,preprintnumbers,amsmath,amssymb,prl,footinbib]{revtex4-1}

\usepackage[titletoc,toc,title]{appendix}
\usepackage[english]{babel}
\usepackage[hidelinks]{hyperref}
\usepackage[normalem]{ulem}
\usepackage{adjustbox}
\usepackage{amsfonts,amsmath,amssymb}
\usepackage{bm}
\usepackage{enumitem}
\usepackage{epsfig}
\usepackage{cancel}
\usepackage{centernot}
\usepackage{color}
\usepackage{comment}
\usepackage{contour}
\usepackage{flushend}
\usepackage{graphics}
\usepackage{graphicx}
\usepackage{epsfig}
\usepackage{mathrsfs}
\usepackage{mdframed}
\usepackage{mathtools}
\usepackage{pifont}
\usepackage{relsize}
\usepackage{shadowtext}
\usepackage{slashed}
\usepackage{soul}
\usepackage{subfigure}
\usepackage{tcolorbox}
\usepackage{textgreek}
\usepackage{times}
\usepackage{titlesec}
\usepackage{titletoc}
\usepackage{verbatim}
\usepackage{xcolor}
\usepackage{xfrac}

\contourlength{0.2em}

\definecolor{myred}{RGB}{123, 28, 20}
\definecolor{zima_blue}{HTML}{1393C1}
\hypersetup{setpagesize=false,bookmarksnumbered=true,bookmarksopen=true,colorlinks=true,linkcolor=zima_blue,urlcolor=zima_blue,citecolor=zima_blue,linktocpage=false}

\begin{document}

\title{Opening the Window of Ultra-Light PBHs by Exorcising the Poltergeist }

\author{Yann Gouttenoire}
\email{yann.gouttenoire@gmail.com}
\affiliation{Institut d’Astrophysique de Paris (IAP), CNRS, Sorbonne Universit\'e, FR-75014, France}
\affiliation{PRISMA$^{++}$ Cluster of Excellence $\&$ Mainz Institute for Theoretical Physics, Johannes Gutenberg University, 55099 Mainz, Germany}
\author{Nicholas Leister}
\email{nleister@uni-mainz.de}
\affiliation{PRISMA$^{++}$ Cluster of Excellence $\&$ Mainz Institute for Theoretical Physics, Johannes Gutenberg University, 55099 Mainz, Germany}
\author{Pedro Schwaller}
\email{pedro.schwaller@uni-mainz.de}
\affiliation{PRISMA$^{++}$ Cluster of Excellence $\&$ Mainz Institute for Theoretical Physics, Johannes Gutenberg University, 55099 Mainz, Germany}

\begin{abstract}
The hot Big Bang may have emerged from evaporation of primordial black holes (PBHs) lighter than $10^9$~g. Standard monochromatic treatments predict nearly simultaneous evaporation, abrupt reheating, and a large \textit{Poltergeist} scalar-induced gravitational wave signal. We confront this expectation with the irreducible collapse mass tail predicted by general relativity, $df_{\rm PBH}/d\ln M\propto M^{3.78}$, which smooths reheating, suppresses the signal by orders of magnitude, and reopens the ultra-light PBH window.

\vspace*{5pt}
\noindent
\textit{Published version:~\href{https://doi.org/10.1103/bgv1-lpq7}{Phys. Rev. Lett. 137, 181002}}

\end{abstract}

\preprint{MITP-26-023}

\maketitle

\textit{\textbf{Introduction.}}  
Primordial black holes provide a direct probe of otherwise inaccessible epochs in the early Universe. Heavy PBHs remain viable dark-matter candidates in parts of parameter space~\cite{Frampton:2010sw,Carr:2016drx,Inomata:2017okj,Green:2020jor}, while PBHs lighter than roughly $10^9\,{\rm g}$ evaporate before Big Bang nucleosynthesis (BBN)~\cite{Kohri:1999ex,Carr:2009jm}. If sufficiently abundant, such ultra-light PBHs dominate the energy density before they evaporate and thereby provide a minimal Standard-Model reheating mechanism~\cite{Carr:1976zz}. They can also produce dark matter through Hawking radiation~\cite{Fujita:2014hha,Lennon:2017tqq,Morrison:2018xla,Franciolini:2026fdv} or participate in mechanisms for baryogenesis~\cite{Toussaint:1978br,Turner:1979bt,Barrow:1990he,Majumdar:1995yr,Dolgov:2000ht,Baumann:2007yr}.

Such a cosmological history leaves several gravitational-wave (GW) signatures. The most promising signals are scalar-induced gravitational waves (SIGWs), generated at second order in cosmological perturbation theory~\cite{Matarrese:1993zf,Matarrese:1997ay,Ananda:2006af,Baumann:2007zm,Saito:2008jc,Espinosa:2018eve,Kohri:2018awv,Domenech:2021ztg}. For PBH domination, the best-known source is the Poltergeist mechanism~\cite{Inomata:2019ivs,Inomata:2020lmk,Papanikolaou:2020qtd,Domenech:2020ssp,Domenech:2021wkk,Bhaumik:2022pil,Gross:2024wkl,Gross:2025hia,Balaji:2024hpu,Domenech:2024cjn,Domenech:2025ffb,Inomata:2025wiv,Papanikolaou:2025ddc,Domenech:2023fuz,Domenech:2024kmh}: scalar modes freeze during matter domination and begin to oscillate once evaporation restores radiation domination, while efficiently sourcing tensor modes.

The predicted amplitude is highly sensitive to the duration of reheating. The studies cited above~\cite{Inomata:2020lmk,Papanikolaou:2020qtd,Domenech:2020ssp,Domenech:2021wkk,Bhaumik:2022pil,Gross:2024wkl,Gross:2025hia,Balaji:2024hpu,Domenech:2024cjn,Domenech:2025ffb,Inomata:2025wiv,Papanikolaou:2025ddc,Domenech:2023fuz,Domenech:2024kmh} assumed a monochromatic PBH population, for which evaporation is nearly instantaneous and the Poltergeist enhancement is maximized. However, PBHs formed from the collapse of super-horizon overdensities~\cite{Zeldovich:1967lct,Hawking:1971ei,Carr:1974nx,Carr:1975qj} do not have identical masses. Critical collapse implies a universal low-mass tail in the PBH mass function~\cite{Choptuik:1992jv,Evans:1994pj,Niemeyer:1997mt,Niemeyer:1999ak}, independent of the detailed shape of the primordial curvature power spectrum. Earlier works already showed that broadening the mass distribution reduces the Poltergeist amplitude~\cite{Inomata:2020lmk}. Here we emphasize that a nonzero width is unavoidable critical collapse in General Relativity alone enforces a universal low-mass tail, and thereby sets an upper envelope on the resulting GW signal.

Besides the Poltergeist signal generated at reheating, the total SIGW spectrum also receives contributions from PBH formation, from the preceding radiation era, and from the PBH-dominated era itself. These additional channels become important once the Poltergeist enhancement is suppressed. The total spectrum then becomes comparable to that of a generic early matter-dominated era, rather than parametrically larger. As a result, both the reach of future GW experiments and the BBN bound are relaxed.

In this \textit{Letter}, we present and summarize the main results, while detailed derivations are deferred to~\cite{Gouttenoire:2026pmp}.

\textit{\textbf{Irreducible PBH mass spread.}} %
We consider PBHs formed during radiation domination from overdensities that collapse when they re-enter the Hubble horizon (for reviews see Refs.~\cite{Carr:1975qj,Escriva:2022duf}). Close to the threshold, numerical simulations, solely relying on general relativity and spherical symmetry, give the universal critical scaling law~\cite{Choptuik:1992jv,Evans:1994pj,Niemeyer:1997mt,Niemeyer:1999ak}
\begin{equation}
\label{eq:critical_scaling_letter_v3}
\text{Choptuik's law:}\quad M_{\rm PBH}=\kappa M_H(\delta_m-\delta_c)^{\gamma_{\rm M}},
\end{equation}
where $M_H$ is the horizon mass, $\delta_c$ is the collapse threshold, $\kappa=\mathcal{O}(1)$, and $\gamma_{\rm M}\simeq0.36$ for radiation domination~\cite{Choptuik:1992jv,Evans:1994pj,Young:2019yug,Ianniccari:2024ltb}. Since $\delta_m$ is continuous, Eq.~\eqref{eq:critical_scaling_letter_v3} maps even a sharply peaked primordial spectrum into an extended PBH mass function~\cite{Kuhnel:2015vtw}.

Press--Schechter and peak-theory computations make this statement explicit~\cite{Press:1973iz,Bardeen:1985tr,Bond:1990iw,Young:2014ana,Young:2019yug,Gow:2020bzo,Karam:2022nym}. We define the normalized PBH mass function in terms of the energy density fraction at formation $\beta_f$ as
\begin{equation}
\label{eq:psi_def_letter_v3}
\psi_f(M)
\equiv
\frac{1}{\beta_f}\frac{d\beta_f}{d\ln M},
\qquad
\int d\ln M\,\psi_f(M)=1 .
\end{equation}
Critical scaling maps the probability distribution of the linear density
contrast at horizon crossing into $\psi_f$. For both Press--Schechter and peak
theory this gives
\begin{equation}
\label{eq:PS_mass_function_letter_v3}
\psi_f(M)\propto
(M/M_H)^{1+1/\gamma_{\rm M}}P_H[\delta_m(M)],
\end{equation}
with $\delta_m(M)$ obtained from
Eq.~\eqref{eq:critical_scaling_letter_v3}. The quantity $P_H$ denotes the
probability distribution of the smoothed density contrast at horizon crossing. 
Press--Schechter and peak theory differ only in how this distribution is
computed. The prefactor in Eq.~\eqref{eq:PS_mass_function_letter_v3} is instead
the common Jacobian from $\delta_m$ to $M$ and fixes the universal low-mass tail,
\begin{equation}
\label{eq:psi_tail_letter_v3}
\psi_f(M)\propto M^{1+1/\gamma_{\rm M}}\simeq M^{3.78}.
\end{equation}
The high-mass tail depends on the statistics, amplitude and shape of the primordial perturbations. For example, it is often parametrized as $\psi_f\propto M^{1+1/\gamma_{\rm M}}\exp[-c_1(M/\langle M\rangle)^{c_2}]$, with $c_1$ and $c_2$ model dependent~\cite{Gouttenoire:2026pmp}. To remain agnostic, and to obtain the largest possible GW signal, we discard this tail and use the sharpest distribution compatible with Eq.~\eqref{eq:psi_tail_letter_v3},
\begin{equation}
\label{eq:psi_sharp_letter_v3}
\psi_f^{\rm sharp}(M)=
\frac{1+\gamma_{\rm M}}{\gamma_{\rm M}}
\left(\frac{M}{M_{\rm cut}}\right)^{1+1/\gamma_{\rm M}}
\Theta(M_{\rm cut}-M) .
\end{equation}
The mean mass is then $\langle M\rangle=(1+\gamma_{\rm M})M_{\rm cut}/(1+2\gamma_{\rm M})$. Any smoother ultraviolet tail prolongs evaporation and further suppresses the GW signal. The mass function in Eq.~\eqref{eq:psi_sharp_letter_v3} should therefore be understood as a conservative upper limit on the linear SIGW amplitude. 

\textit{\textbf{PBH reheating.}} 
On scales larger than their mean separation, PBHs are described as a pressureless fluid with an additional evaporation term. For a single PBH,
\begin{equation}
\label{eq:PBH_decay_rate_letter_v3}
M(t)=M_f\left(1-\frac{t}{t_{\rm eva}}\right)^{1/3},
\qquad
 t_{\rm eva}=\frac{M_f^3}{3A M_{\rm pl}^4},
\end{equation}
where $A=(\pi\mathscr{G}/480)g_{H\star}$ includes greybody factors and the number of emitted degrees of freedom~\cite{Hawking:1974sw,Page:1976df,Hooper:2019gtx}. If the initial fraction $\beta_f$ exceeds a critical value $\beta_c$, PBHs dominate before they evaporate. 
For an extended mass function, all background quantities are averaged over $\psi_f$, yielding 
the homogeneous evolution
\begin{align}
\label{eq:background_averaged_letter_v3}
\langle\dot\rho_{\rm PBH}\rangle+3H\langle\rho_{\rm PBH}\rangle&=-\langle\Gamma\rho_{\rm PBH}\rangle,
\\
\langle\dot\rho_{\rm r}\rangle+4H\langle\rho_{\rm r}\rangle&=\langle\Gamma\rho_{\rm PBH}\rangle.
\end{align}
Thus the decay rate is not the rate of a single PBH but the convolution of Eq.~\eqref{eq:PBH_decay_rate_letter_v3} with $\psi_f$. Different masses disappear at different times, smoothening the matter-to-radiation transition. The relevant comoving scales are the evaporation scale $k_{\rm eva}=a_{\rm eva}H_{\rm eva}$, the equality scale $k_{\rm eq}=a_{\rm eq}H_{\rm eq}$ at the onset of PBH domination, and the discreteness scale $k_{\rm PBH}$ set by the mean inter-PBH separation. These scales determine which modes enter during the first radiation era, during PBH domination, 
or after evaporation.

This smoothing is most transparently seen in the Newtonian potential, sourced by two scalar seeds: the adiabatic mode $\Phi$ inherited from inflation, and the PBH isocurvature seed
\begin{equation}
\label{eq:isocurvature_def_letter}
S \equiv \delta_{\rm PBH} - \frac{3}{4}\,\delta_{\rm rad},
\end{equation}
which measures the local imbalance between PBH and radiation densities. Being discrete objects, PBHs carry unavoidable Poisson fluctuations on top of the smooth radiation background. The potential released at the onset of the final radiation era reads
\begin{equation}
\label{eq:Phi_osc_letter_v3}
\begin{aligned}
\Phi_{\rm osc}(k)
=
\Big[
T_\Phi^{\rm (eMD)}(k)\,\Phi(k,0)
+T_S^{\rm (eMD)}(k)\,S(k,0)
\Big]\mathcal{S}_\Phi(k) ,
\end{aligned}
\end{equation}
where $T_X^{\rm (eMD)}(k)$, with $X\in\{\Phi,S\}$, is the transfer function propagating each seed through the radiation-to-matter transition and freezing sub-horizon modes during PBH domination, while $\mathcal{S}_\Phi$ captures the decay of the potential during evaporation. Both approach constants on large scales but suppress modes that entered deep in the first radiation era. A monochromatic mass function gives the standard scaling $\mathcal{S}_\Phi\propto k^{-1/3}$~\cite{Inomata:2020lmk,Domenech:2020ssp}, whereas Eq.~\eqref{eq:psi_sharp_letter_v3} ends evaporation in a universal cusp where the remaining PBH fraction scales as $(\eta_{\rm cut}-\eta)^{4/3}$. A mode of wavenumber $k$ resumes oscillating once the Hubble rate drops below $k$, a conformal interval $\eta_{\rm cut}-\eta_{\rm osc}\propto k^{-1}$ before the cusp, giving
\begin{equation}
\label{eq:suppression_letter_v3}
\mathcal{S}_\Phi(k)\simeq
\kappa_0\left(\frac{k_{\rm eva}}{k}\right)^{4/3},
\qquad \kappa_0\simeq2.3,
\end{equation}
for $\gamma_{\rm M}\simeq0.36$. Since $\Omega_{\rm GW}\propto\mathcal{S}_\Phi^4$, the change from $k^{-1/3}$ to $k^{-4/3}$ suppresses the Poltergeist contribution by orders of magnitude.

\begin{figure}[ht]
\centering
\includegraphics[width=0.47\textwidth]{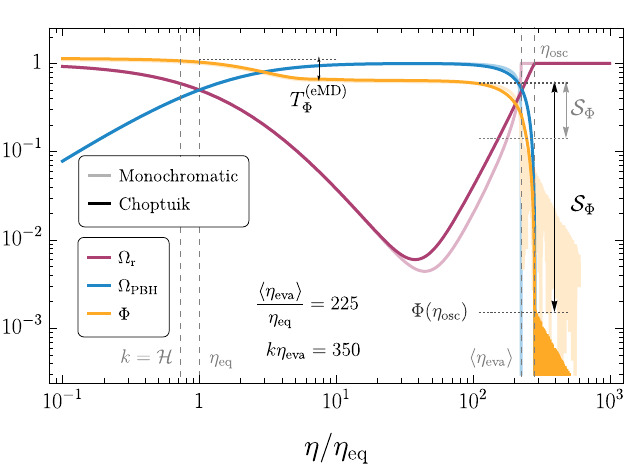}
\caption{\label{fig:PBHOmega} \it Evolution of $\Omega_{\rm PBH}$ (blue), $\Omega_{\rm r}$ (purple), and the Newtonian potential (orange) during PBH domination. The transitions are described by $T$ and the evaporation suppression factor $\mathcal{S}_\Phi$. Faded curves assume a monochromatic PBH mass distribution; solid curves include Choptuik broadening.
}
\end{figure}

\textit{\textbf{Scalar-induced gravitational waves.}} %
The relevant sequence of epochs is
\begin{equation}
\label{eq:eras_letter_v3}
{\rm formation}\longrightarrow {\rm RD}_1\longrightarrow {\rm eMD}\longrightarrow {\rm RD}_2 .
\end{equation}
We label by ${\rm RD}_1$ and ${\rm RD}_2$ the radiation eras before PBH domination and after Hawking reheating. Both are radiation dominated but their scalar sources differ. ${\rm RD}_1$ contains the adiabatic radiation mode and the Poisson isocurvature of the PBH gas, while the intervening eMD phase freezes the Newtonian potential before Hawking evaporation restores the radiation fluid in ${\rm RD}_2$. We decompose the present SIGW signal by emission epoch (superscript) and scalar source $X$,
\begin{align}
\label{eq:GW_total_letter_v3}
\Omega_{\rm GW}
&\simeq
\Omega_{\rm GW}^{\rm (form)}[\Phi]
\\&+
\sum_{X=\Phi,S}
\left[
\Omega_{\rm GW}^{\rm (RD_1)}[X]
+
\Omega_{\rm GW}^{\rm (eMD)}[X]
+
\Omega_{\rm GW}^{\rm (RD_2)}[X]
\right] .\nonumber
\end{align}
The last term contains the Poltergeist background, but Eq.~\eqref{eq:GW_total_letter_v3} shows that it is only one component of the total PBH signal. The adiabatic seed follows the usual nearly scale-invariant spectrum, $\mathcal{P}_\Phi\simeq (2/3)^2\mathcal{A}_s(k/k_{\rm CMB})^{n_s-1}$~\cite{Planck:2018vyg}, while the PBH-isocurvature seed arises from number-density fluctuations of the discrete PBH gas. At formation,
$\mathcal{P}_{S,f}(k)=2/(3\pi)\left(k/k_{\rm PBH}\right)^3
\Theta(k_{\rm PBH}-k)$,
with a cutoff at the scale set by the mean comoving PBH separation~\cite{Ali-Haimoud:2018dau,Carr:2018rid,Papanikolaou:2020qtd,Domenech:2020ssp,Gerlach:2025vco}.

All these terms are computed from the standard second-order tensor equation. In terms of the tensor power spectrum, the GW density after oscillation averaging is
\begin{equation}
\label{eq:omega_basic_letter_v3}
\Omega_{\rm GW}(k,\eta)=\frac{1}{12}
\left(\frac{k}{\mathcal{H}}\right)^2
\overline{\mathcal{P}_h(k,\eta)},
\end{equation}
where $\mathcal{P}_h$ is obtained by convolving two scalar transfer functions with the tensor Green function during radiation and matter eras. We neglect mixed terms between statistically independent sources and between widely separated production epochs. Constructive interference can enhance the total signal by at most a factor of $2$. The ${\rm RD}_2$ contribution splits into an irreducible radiation-era background for modes entering after evaporation, and the Poltergeist component for modes already inside the horizon. The latter has the standard radiation-era kernel, but with each scalar leg multiplied by $T_X^{\rm (eMD)}\mathcal{S}_\Phi$. Schematically,
\begin{widetext}
\begin{equation}
\label{eq:Omega_RD2_schematic_v3}
\Omega_{\rm GW}^{\rm (RD_2,Polt.)}[X]
\propto
\left(\frac{k}{k_{\rm eva}}\right)^8
\int d u\,d v\,
\mathcal{K}(u,v)\,\overline{\mathcal{I}_{\rm osc}^2}(u,v)\,
\mathcal{S}_\Phi^2(uk)\mathcal{S}_\Phi^2(vk)
T_X^2(uk)T_X^2(vk)
\mathcal{P}_X(uk)\mathcal{P}_X(vk),
\end{equation}
\end{widetext}
where $X=\{\Phi,S\}$ denotes adiabatic and isocurvature initial conditions,
$\mathcal{K}$ denotes the $\Phi\Phi\xrightarrow{} h$ kinematic scattering kernel
\cite{Domenech:2020ssp,Domenech:2024wao} and $\overline{\mathcal{I}_{\rm osc}^2}(u,v)$ the oscillation averaged GW kernel. The explicit factor
$(k/k_{\rm eva})^8$ is the origin of the large Poltergeist enhancement for monochromatic PBHs. The integral over $u$ and $v$ is dominated by the resonance occurring when the two scalar mode frequencies $uk/\sqrt{3}$ and $vk/\sqrt{3}$ together
match the tensor frequency $k$.
Along the resonant contribution of the kernel the averaged time integral contributes a factor $k_{\rm eva}/k$. Thus, using
$\mathcal{S}_\Phi\propto (k/k_{\rm eva})^{-4/3}$ from
Eq.~\eqref{eq:suppression_letter_v3}, one obtains
\begin{equation}
\label{eq:polt_power_counting_v3}
\left(\frac{k}{k_{\rm eva}}\right)^8
\left(\frac{k_{\rm eva}}{k}\right)
\left[\mathcal{S}_\Phi(k)\right]^4
\propto
\left(\frac{k}{k_{\rm eva}}\right)^{5/3}.
\end{equation}
The resonant Poltergeist contribution therefore scales as
\begin{align}
\label{eq:polt_peak_letter_v3}
&\Omega_{\rm GW}^{\rm (RD_2,Polt.)}[X]
\propto
\mathcal{P}_X^2(k)
\left(\frac{k}{k_{\rm eva}}\right)^{5/3}
\left[T_X^{\rm (eMD)}(k)\right]^4\end{align}
and yield for $X=\{\Phi,S\}$:
\begin{align}
&\!\!\!\!\Omega_{\rm GW}^{\rm (RD_2,Polt.)}[\Phi]
\simeq 
\mathcal{A}_{\mathcal{R}}^2\!
\left(\frac{k}{k_{\rm eva}}\right)^{\!\frac{5}{3}}\!\!\!\!
\frac{
7.7\times 10^{-3}
}{
\bigl[1+ 0.24 \,(k/k_{\rm eq})^{0.8}\bigr]^9
},\!\!\label{eq:GW_RD_2_polt_Phi}\\
&\!\!\!\!\!\!\Omega_{\rm GW}^{\rm (RD_2,Polt.)}[S]\simeq\!
\left(\frac{k}{k_{\rm \mathsmaller{PBH}}}\right)^{\!6}\!\!
\left(\frac{k}{k_{\rm eva}}\right)^{\!\frac{5}{3}}
\!\!\!\!\!
\frac{
1.5\times 10^{-6}
}{
\left[
1+0.38\left(k/k_{\rm eq}\right)
\right]^8
},
\end{align}
where we assume scalar spectral index $n_s \simeq 1 $ in Eq.~\eqref{eq:GW_RD_2_polt_Phi}.
For a monochromatic mass function one would instead have
$\mathcal{S}_\Phi\propto (k/k_{\rm eva})^{-1/3}$, giving the much steeper
scaling $(k/k_{\rm eva})^{17/3}$. This replacement is the parametric reason why
the Choptuik mass spread removes the dominance of the Poltergeist peak. The
isocurvature term follows the same counting, but its Poisson spectrum weights it
more strongly toward the largest available scalar momenta.

Besides the reheating contribution, the PBH cosmology contains several
additional SIGW channels~\cite{Gouttenoire:2026pmp}:
\begin{itemize}
    \item \emph{eMD contribution.}
    Before evaporation, the PBH gas drives an early matter-dominated era.
    During this phase the Newtonian potential is approximately constant on
    sub-horizon scales~\cite{Kohri:2018awv,Inomata:2019ivs}, so the tensor source remains active for an extended time. In a pure matter era this gives the familiar growth of the induced GW signal with the scale factor $\Omega_{\rm GW}\propto a$. Modes entering well before equality are suppressed. The eMD signal therefore peaks near $k_{\rm eq}\equiv \mathcal{H}_{\rm eq}$, the comoving Hubble horizon when PBHs start dominating the universe. We find that the eMD contribution is of similar magnitude as the Poltergeist component with the Choptuik mass function:
    \begin{align}\Omega_{\rm GW}^{\rm (eMD)}[X] &\simeq C_X\mathcal P_X^2(k_{\rm UV}) \frac{k_{\rm UV}k}{\mathcal H^2}\frac{0.02\,x_{\rm eva}^{2}} {1+0.01\,x_{\rm eva}^{10/3}},\!\!\!\!\!\!\!\!\!\!\!
    \end{align}
    with $C_{\Phi,S}= \left(81/125,16/70875\right)$ and $x_{\rm eva}\equiv 2k/k_{\rm eva}$. 
    The adiabatic and isocurvature components have been studied in ~\cite{Kohri:2018awv,Inomata:2019zqy,Pearce:2023kxp,Pearce:2025ywc,Fernandez:2023ddy} and \cite{Papanikolaou:2022chm}, respectively. The latter also studies an extended mass distribution but obtain a much larger signal.
    \item \emph{Universal ${\rm RD}_1$ isocurvature contribution.}
    Even if the primordial curvature spectrum is featureless on the relevant
    scales, the discreteness of the PBH population generates Poisson
    fluctuations. These act as an isocurvature source and produce a universal
    SIGW background during the first radiation era~\cite{Domenech:2021and,Lozanov:2023aez,Lozanov:2023knf,Domenech:2023fuz,Domenech:2025ffb}. This channel follows only
    from treating PBHs as localized objects with finite number density, and is
    therefore less model-dependent than the formation signal. Since it is
    produced before PBH domination and evaporation, its present-day amplitude is
    diluted by the entropy release,
    \begin{equation}
        \Omega_{{\rm GW}}^{({\rm RD}_1)}[S]
        \simeq
        \frac{10^{-3}(k_{\rm eq}/k_{\rm PBH})^4}{D^{4/3}}.
    \end{equation}
  Here $D$ denotes the entropy dilution induced by PBH evaporation,
\begin{equation}
    D
    \equiv
    \frac{S(T_{\rm eva})}{S(T_{\rm eva}^{-})}
    \simeq
    \frac{\beta_f}{\beta_c}
    \simeq
    \exp\!\left(\frac{3}{4}N_{\rm MD}\right),
\end{equation}
where $S(T)$ is the comoving entropy and $T_{\rm eva}^{-}$ denotes the
temperature immediately before evaporation. In the absence of PBH domination,
no entropy is injected and $D=1$.

    \item \emph{Irreducible adiabatic radiation-era contribution.}
    Adiabatic curvature perturbations also source SIGW during the standard radiation era, which in our case are $\rm RD_1$ and $\rm RD_2$~\cite{Ananda:2006af,Baumann:2007zm,Kohri:2018awv}. The part generated before PBH domination is diluted by the subsequent entropy
    injection, while the part generated after evaporation is not,
    \begin{equation}
\Omega_{\rm GW}^{(\rm RD_1)}[\Phi]\simeq\frac{\Phi_{\rm CMB}^4}{D^{4/3}},\quad \Omega_{\rm GW}^{(\rm RD_2)}[\Phi]\simeq \Phi_{\rm CMB}^4.
\end{equation}
    These irreducible components are controlled by the small CMB normalized
    scalar amplitude $\Phi_{\rm CMB}^2\sim 10^{-9}$ and therefore remain far below the eMD and reheating
    signals, but they provide a lower bound on the GW background.

    \item \emph{Formation contribution.}
    The enhanced curvature perturbations  $\Phi_{\rm PBH}^2\sim 10^{-2} $ that produce the PBHs source an
    additional SIGW background when they re-enter the horizon
    \cite{Saito:2008jc,Pi:2020otn}. Since it is generated deep in ${\rm RD}_1$, it is also diluted by
    the later eMD epoch,
    \begin{equation}
        \Omega_{{\rm GW}}^{({\rm form})}[\Phi]
        \simeq
        D^{-4/3}\,\Phi_{\rm PBH}^4.
    \end{equation}
\end{itemize}
The SIGW integrals require an ultraviolet prescription for the largest scalar
momenta included in the source. We denote this scale by $k_{\rm UV}$. We distinguish two relevant choices:
\begin{itemize}
    \item \emph{Non-linear cutoff.}
    Our fiducial choice is the scale $k_{\rm NL}$ at which the PBH density
    contrast becomes order unity by the end of eMD. For adiabatic and
    Poisson-isocurvature seeds,
    \begin{equation}
    \label{eq:kNL_letter_v3}
    \begin{aligned}
    k_{\rm NL}^{\rm adia}
    &\simeq
    235\,k_{\rm eva}
    \left(\frac{k_{\rm eva}}{k_{\rm CMB}}\right)^{(1-n_s)/4},
    \\
    k_{\rm NL}^{\rm iso}
    &\simeq
    \left(\frac{675\pi}{8}\right)^{1/7}
    k_{\rm eva}^{4/7} k_{\rm PBH}^{3/7},
    \end{aligned}
    \end{equation}
    up to order-one transfer-function factors. We therefore take
    $k_{\rm UV}=k_{\rm NL}$ as the conservative perturbative cutoff.
    \item \emph{Fluid cutoff.}
    The formal upper limit is the inverse PBH separation, $k_{\rm PBH}$. At this
    scale the coarse-grained fluid description of the PBH gas breaks down.
    Integrating to $k_{\rm PBH}$ yields therefore an upper limit.
\end{itemize}
We therefore use $k_{\rm UV}=k_{\rm NL}$ as our fiducial cutoff. Results with $k_{\rm UV}=k_{\rm PBH}$ should instead be read as an extrapolation
to the smallest scale on which the PBH gas can still be coarse grained. Once
modes become non-linear, additional GW production from halo formation may occur \cite{Fernandez:2023ddy}, which goes beyond the linear SIGW
calculation used here.

\begin{figure}[t]
\centering
\includegraphics[width=0.46\textwidth]{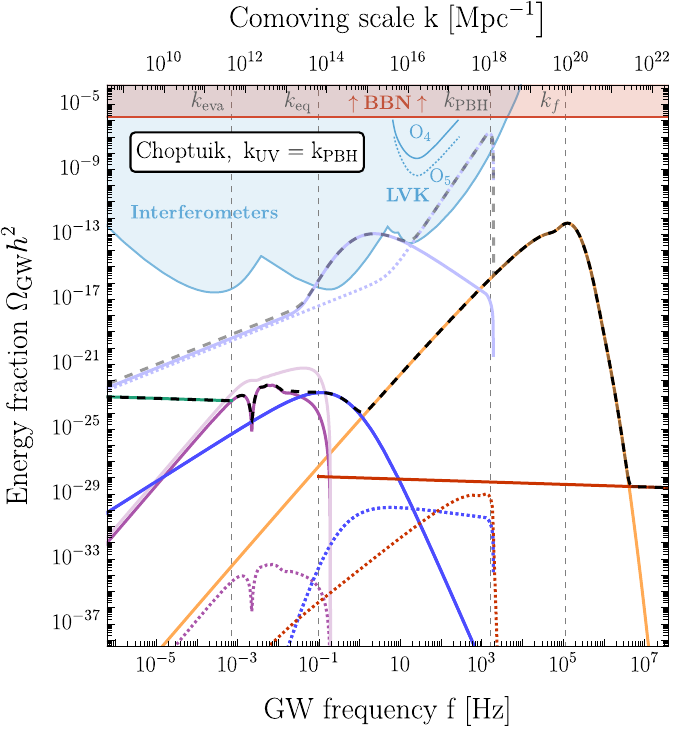}\vspace{0.2cm}
\includegraphics[width=0.46\textwidth]{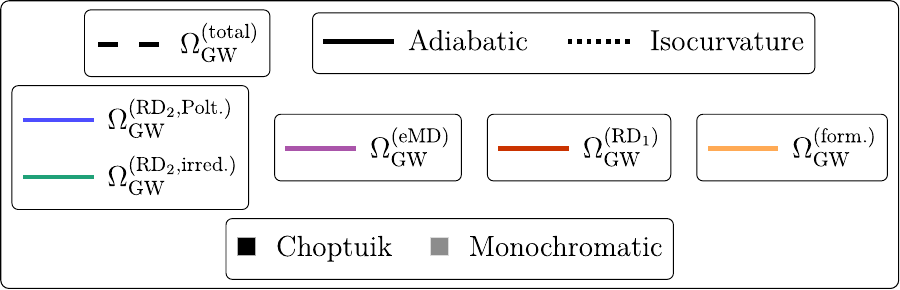}
\caption{\it Channel decomposition of the scalar-induced GW signal from PBH reheating. We show the contributions generated at PBH formation, during the first radiation era, during PBH domination, and at reheating, for $M_{\rm PBH}=10^4,{\rm g}$ and $\beta_f=6\times 10^{-7}$. The spectra are cut off at the PBH interdistance scale, $k_{\rm UV}=k_{\rm PBH}$, and should therefore be interpreted as optimistic upper envelopes.
}
\label{fig:GWSig_letter_v3}
\end{figure}

Fig.~\ref{fig:GWSig_letter_v3} shows the resulting SIGW spectrum as a sum of
physically distinct sources for the most aggressive cut-off $k_{\rm UV}=k_{\rm PBH}$. The Choptuik scaling smoothens the reheating epoch and suppresses the Poltergeist peak. The reheating
signal then becomes comparable to the contribution accumulated during eMD,
while the ${\rm RD}_1$ term provides a smaller irreducible
component. In this respect, apart from the PBH formation component, the SIGW spectrum resembles that of a generic eMD era sourced by an unstable heavy particle with
a constant decay rate.

\begin{figure*}[t] \centering \begin{minipage}[c]{0.6\textwidth} \centering \includegraphics[width=\linewidth]{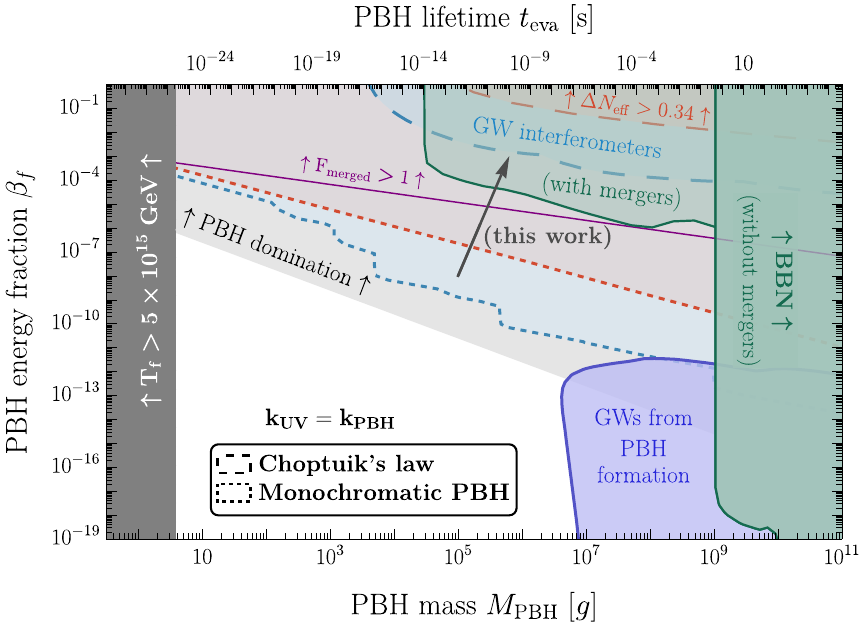} \end{minipage} \hfill \begin{minipage}[c]{0.38\textwidth} \caption{\itshape Exclusion and sensitivity contours in the $(M_{\rm \mathsmaller{PBH}}, \beta_f)$ plane of the ultra-light PBH parameter space. We compare a monochromatic PBH gas (dotted) with one having the Choptuik infrared tail in the mass function (dashed), as predicted by gravitational collapse in general relativity. In the Choptuik case, the smoother reheating transition suppresses the Poltergeist SIGW signal and, in consequence, relaxes the BBN bound on $\Delta N_{\rm eff}$ (red) and the GW interferometer reach (blue), reopening regions of the parameter space that were either excluded or reachable in the monochromatic analysis. Sensitivity curves correspond to the most optimistic combination of planned GW interferometers (SKA, Theia, $\mu$Ares, LISA, AEDGE, BBO, DECIGO, ET, and CE), assuming total astrophysical foreground subtraction.  The figure adopts the optimistic cutoff, $k_{\rm UV}=k_{\rm PBH}$, while for the conservative one, $k_{\rm UV}=k_{\rm NL}$, the interferometer prospects and $\Delta N_{\rm eff}$ constraints vanish.}\label{fig:beta_vs_MPBH_letter_v3} \end{minipage} \end{figure*}

\textit{\textbf{Parameter space.}} %
We scan the plane $(\langle M_{\rm PBH}\rangle,\beta_f)$. PBH domination
requires $\beta_f>\beta_c$, while BBN requires sufficiently early evaporation, i.e. $M_{\rm PBH}\lesssim 10^9~\rm g$.
The GW background is redshifted to today as
\begin{equation}
\label{eq:redshift_letter_v3}
\Omega_{{\rm GW},0}h^2\simeq
0.39\left(\frac{g_\star(T_{\rm eva})}{106.75}\right)^{\!-1/3}\!\!
\Omega_{r,0}h^2\,\Omega_{\rm GW}(T_{\rm eva}),
\end{equation}
and contributes to extra radiation $\Delta N_{\rm eff}$.
Fig.~\ref{fig:beta_vs_MPBH_letter_v3} shows the probing potential of ultra-light PBHs including all SIGW components. For a monochromatic mass function, the signal, dominated by the Poltergeist component, can violate
$\Delta N_{\rm eff}<0.34$. Instead, with the Choptuik infrared tail of the PBH mass distribution, the BBN bound and the reach by GW observatories~\cite{Crowder:2005nr,Punturo:2010zz,LIGOScientific:2014pky,Janssen:2014dka,Reitze:2019iox,AEDGE:2019nxb,Badurina:2019hst,Sesana:2019vho,Garcia-Bellido:2021zgu,KAGRA:2021kbb,Blas:2021mqw,LISACosmologyWorkingGroup:2022jok,NANOGrav:2023hvm} are entirely relaxed even with the most aggressive cut-off  $k_{\rm UV}=k_{\rm PBH}$.

Hawking gravitons and GWs from PBH mergers peak at much higher frequencies and give 
subdominant contributions to $\Delta N_{\rm eff}$~\cite{Hooper:2019gtx,Zagorac:2019ekv,Cheek:2022mmy,Gross:2024wkl}. 
A detailed analysis of these signals will be presented in a companion 
paper~\cite{YannNicoPedro}. For reference, in Fig.~\ref{fig:beta_vs_MPBH_letter_v3} we also display in purple the merger contour $F_{\rm merged}>1$, which is estimated by
comparing $t_{\rm eva}$ in Eq.~\eqref{eq:PBH_decay_rate_letter_v3} with the maximal inspiral time given by Peters formulae
\cite{Peters:1964zz}. Above it, repeated mergers modify the PBH mass function
before evaporation, delaying evaporation and strengthening the BBN bound in green~\cite{Holst:2024ubt}  (see also~\cite{Aljazaeri:2025ftv}).

\textit{\textbf{Conclusions.}} 
The large Poltergeist background predicted for a monochromatic PBH population~\cite{Inomata:2020lmk,Papanikolaou:2020qtd,Domenech:2020ssp,Domenech:2021wkk,Bhaumik:2022pil,Gross:2024wkl,Gross:2025hia,Balaji:2024hpu,Domenech:2024cjn,Domenech:2025ffb,Inomata:2025wiv,Papanikolaou:2025ddc,Domenech:2023fuz,Domenech:2024kmh} is not a generic prediction of PBH reheating. 
Critical collapse enforces a universal infrared tail in the PBH mass function. Even the narrowest distribution consistent with this scaling changes the evaporation suppression factor from $\mathcal{S}_\Phi\propto k^{-1/3}$ to $\mathcal{S}_\Phi\propto k^{-4/3}$, suppressing the reheating SIGW signal by orders of magnitude. Within numerical uncertainty or for other equations of state, varying $\gamma_{\rm M}$ shifts only the order-one prefactor of $\mathcal{S}_\Phi$ and preserves its strong high-$k$ suppression~\cite{Gouttenoire:2026pmp}. The infrared scaling law in Eq.~\eqref{eq:psi_tail_letter_v3} holds in any scenario where PBHs form from the gravitational collapse of a statistical
distribution of curvature peaks, and our conclusions therefore apply to a broad class of PBH formation mechanisms.

Consequently, future GW experiments are unlikely to probe ultra-light PBH reheating through GWs, and the BBN bound from GW overproduction is substantially relaxed, reopening the parameter space for PBH-dominated cosmologies.

Our treatment is conservative in two respects. First, the sharp cutoff in Eq.~\eqref{eq:psi_sharp_letter_v3} minimises the smoothing of the reheating transition allowed by {Choptuik's} critical scaling, Eq.~\eqref{eq:critical_scaling_letter_v3}. Realistic high-mass tails would only further suppress the signal. Second, Fig.~\ref{fig:beta_vs_MPBH_letter_v3} assumes the aggressive cutoff $k_{\rm PBH}$. Our claims become even stronger when the more conservative cutoff $k_{\rm NL}$ is used.

\textit{\textbf{Acknowledgments.}}  
The authors thank Amirah Aljazaeri, Marco Calzà, Guillem Domènech, Gabriele Franciolini, Keisuke Inomata, Antonio Iovino, Riccardo Maule, Gabriele Perna, Jan Tränkle, Ville Vaskonen, and Hardi Veermäe for useful discussions.
The authors acknowledge support by the Cluster of Excellence ``PRISMA$^{++}$'' funded by the German Research Foundation (DFG) within the German Excellence Strategy (Project No. 390831469). NL is grateful to the German Academic Scholarship Foundation for the award of a PhD fellowship. PS would like to thank the CERN TH Department for hospitality during completion of this work. 

\textit{\textbf{Note added.}}
Shortly after this work appeared on the arXiv, Ref.~\cite{He:2026neq} independently reported the same scaling as Eq.~\eqref{eq:suppression_letter_v3}.

\bibliographystyle{apsrev4-1}
\bibliography{Letter}

\end{document}